\documentclass[pre,aps,showpacs,eqsecnum,twocolumn,floats]{revtex4}
\usepackage{psfig}
\begin{document}
\title{ Brownian Motion in a Classical Ideal Gas: 
a Microscopic Approach to Langevin's Equation}
\author{Rangan Lahiri}
\thanks{deceased}
\affiliation{Department of Physics, Indian Institute of Science,
Bangalore, India.}
\author{Arvind}
\email{xarvind@andrew.cmu.edu}
\thanks{Permanent Address: Department of Physics, 
Guru Nanak Dev University, Amritsar,143 005, India.}
\affiliation{Department of Physics, Carnegie
Mellon University, Pittsburgh PA 15213.}
\author{Anirban Sain}
\email{asain@onsager.uwaterloo.ca}
\affiliation{Department of Physics, 
University of Waterloo, Ontario, Canada N2L3G1.} 
\begin{abstract}
We present an insightful ``derivation'' of the Langevin
equation and the fluctuation dissipation theorem in the
specific context of a  heavier particle moving through an
ideal gas of much lighter particles.  The Newton's Law of
motion ($m{\ddot x}=F$) for the heavy particle reduces to a
Langevin equation (valid on a coarser time scale) with the
assumption that the lighter gas particles follow a Boltzmann
velocity distribution.  Starting from the kinematics of the
random collisions we show that (1) the average force
$\langle F\rangle \propto -{\dot x}\/$ and (2) the
correlation function of the fluctuating force $\eta=
F-\langle F\rangle\/$ is related to the strength of the
average force. 
\end{abstract}
\pacs{5:40}
\maketitle
\section{Introduction}
In recent years, many experiments on single large molecules
like colloidal spheres and bio-polymers, have been reported
\cite{expt}.  Many of these experiments focus on the
kinematics of motion of these large molecules through typical
fluid environments, at room temperature.  Most theoretical
explanations of these experiments~\cite{expt,single-mol-thry}
neglect the intrinsic correlations of the surrounding fluid
and treat the fluid as a bath, generating friction and
uncorrelated noise on the large molecule, exactly  the way a
Brownian particle is treated.

This approach may fail if the molecule is large or if
it is close to a hard wall such that the displaced fluid due
to its motion does not return to equilibrium immediately.
In that case one may have to solve
the Navier-Stokes equation to correctly incorporate the
dynamics of the fluid, for example while deriving the
Stokes' law~\cite{stokes}. Further, extended macromolecules
of the polymer variety, where one part of the molecule
disturbs the fluid which in turn affects the other parts of
the molecule (the so called hydrodynamic effect
\cite{doi-edwards}) require yet another kind of analysis. It
is however worthwhile and desirable to gain more physical
insight and understanding into the simple yet intriguing
situation of the Brownian motion, for its own sake and
also to come up with  meaningful models for more
complicated situations.  We revisit the problem of Brownian
motion through an ideal gas in this work.

When a particle, substantially heavier than the gas particles,
is injected into the gas, it is randomly kicked around by 
the gas molecules. However, due to the heavier nature of this
particle (usually called the Brownian particle) it takes a
bunch of small kicks by the gas molecules, over a time interval,
to move the Brownian particle appreciably. Thus there occurs a
natural separation of time scales for the motion of this heavier 
particle and the gas molecules~\cite{ein-brown,Lngvneqn}. 
The motion of a Brownian particle, on a coarse grained time
scale  is usually described by the Langevin
equation~\cite{Lngvneqn,chandra},
\begin{equation}
M \ddot{x} + \Gamma \dot{x} = \eta .
\label{langavin-1}
\end{equation}
where  $x\/$ is the position of the Brownian particle.  The
overall effect of collisions is modeled by introducing the
damping coefficient ${\Gamma}$ and a random force  $\eta\/$ 
with zero mean. This is a physically motivated
phenomenological equation where the effect of collisions is
neatly separated into velocity dependent friction and a
random force, with a nontrivial relationship between the two,
the fluctuation dissipation theorem(FDT)~\cite{fdt}.

In the general context of many body systems the Langevin
equation provides a description of the slow degrees of
freedom, while treating the fast modes as noise acting on
the slower ones. Apart from Brownian motion, thermally
activated barrier crossing phenomena~\cite{hanngi},
polymer dynamics in a solvent fluid~\cite{doi-edwards} and
epitaxial growth of a surface profile due to random
deposition~\cite{krug} are  
a few examples of such a situation and the description
assumes that a distinct natural  separation of time-scales
exists between the fast and the slow degrees
of freedom.  The non-triviality of the Langevin description
lies in splitting the effect of the fast degrees of freedom
into a systematic part (friction) and a random noise obeying
FDT.  Mostly, the Langevin equation is motivated from a
phenomenological point of view. 
Every derivation of the Langevin equation requires certain
assumptions; Mori and Zwanzig have given a formal
prescription (projection operator method \cite{garbert}) to
derive the Langevin equation starting from a microscopic
picture. Given a Langevin equation, Kubo's method
\cite{kubo,holland-FDT} can be used to derive FDT with
certain assumptions.
Assuming the motion of the Brownian particle through an
ideal gas to be a jump
Markov process, Gillespie has derived the Langevin 
equation~\cite{Gillespie}. A more
physical approach to derive Langevin equation and FDT,
though in one dimension, has also been reported \cite{AJP2}.

In this paper we present a new approach to 
derive the Langevin equation for a Brownian
particle moving in an ideal gas at a fixed temperature,
using a combination of microscopic and statistical ideas.
The work we present here differs from others mainly in the
way we derive the FDT and that we consider two different
limits of the ideal gas, namely, small mean free path and
large mean free path.

The fact that for a moving Brownian  particle the  relative
velocity of the gas particles approaching it from the front
is higher than the ones approaching from behind, causes
retardation leading to net damping.  So the collisional
force, though random in nature has a finite average $
\langle F\rangle\, (= -\Gamma \dot{x}\,$, in
Eqn.~(\ref{langavin-1})).  If we separate out this time
independent average from the total force, what remains
(i.\thinspace e, $F-\langle F\rangle$) is a time varying
random force $\eta\/$.  For the Langevin description to be
valid the probability distribution of $\eta \/$ must
satisfy FDT.

We explicitly show this for
two different limits of the ideal gas, namely the one with
high collision rate and the other with negligible collision
rate.

The material in this paper is arranged as follows: In
section~\ref{gas-1} we calculate the damping constant  and
derive the fluctuation dissipation theorem for a gas with a
low collision rate.  The gas molecules in this case do not
suffer too many collisions among themselves in  the time
taken by the Brownian particle to gain a sufficient amount of
momentum.  Section~\ref{gas-2} deals with similar questions
for an ideal gas with a high collision rate and in this case
the gas particles can equilibrate in a short time compared
to the time taken by the Brownian particle to gain momentum.
Section~\ref{conc} contains a discussion of the results and
some conclusions.  In appendix~\ref{sphere} we discuss the
case when the Brownian particle is assumed  to be a sphere
instead of a disk and appendix~\ref{math} has some useful
mathematical formulae 
\section{Brownian motion in an Ideal gas with mean free path
$\lambda \rightarrow \infty$} 
\label{gas-1}
Consider a Brownian particle `B' moving in a gas along the
$x\/$ direction with a velocity $u_0\/$. To find the
friction coefficient $\Gamma\/$, we have to
calculate the net force on B due to collisions with
the gas molecules. Consider the simplest
possible geometry for B, that of a plate moving  with
a velocity perpendicular to its plane (In
appendix~\ref{sphere}, 
the calculation is  extended to a spherical
particle). 
It is easier to capture the essentials of the
calculation in this case without getting lost in the
geometrical factors. The easiest way to arrive at the force
is to treat all collisions in the rest frame of the
Brownian particle. If a molecule comes in with a velocity
${\bf v}\/$ then from the conservation of energy and
momentum, it follows that it is reflected with 
a velocity $- \kappa
{\bf v}\/$ 
\begin{equation}
\kappa = \frac{M-m}{M+m}
\label{kappa}
\end{equation}
$M\/$ and $m\/$ being the masses of the Brownian particle 
and of the gas molecule respectively.
For $M \gg m$, the momentum transfered to B per collision is then 
\begin{equation}
\delta {\bf p} = (1+\kappa)\/m\/{\bf v}
\label{momentum/col}
\end{equation}

\begin{figure}
\unitlength=.8mm
\begin{picture}(80,80)(-20,0)
\boldmath
\put(0,0){\psfig{figure=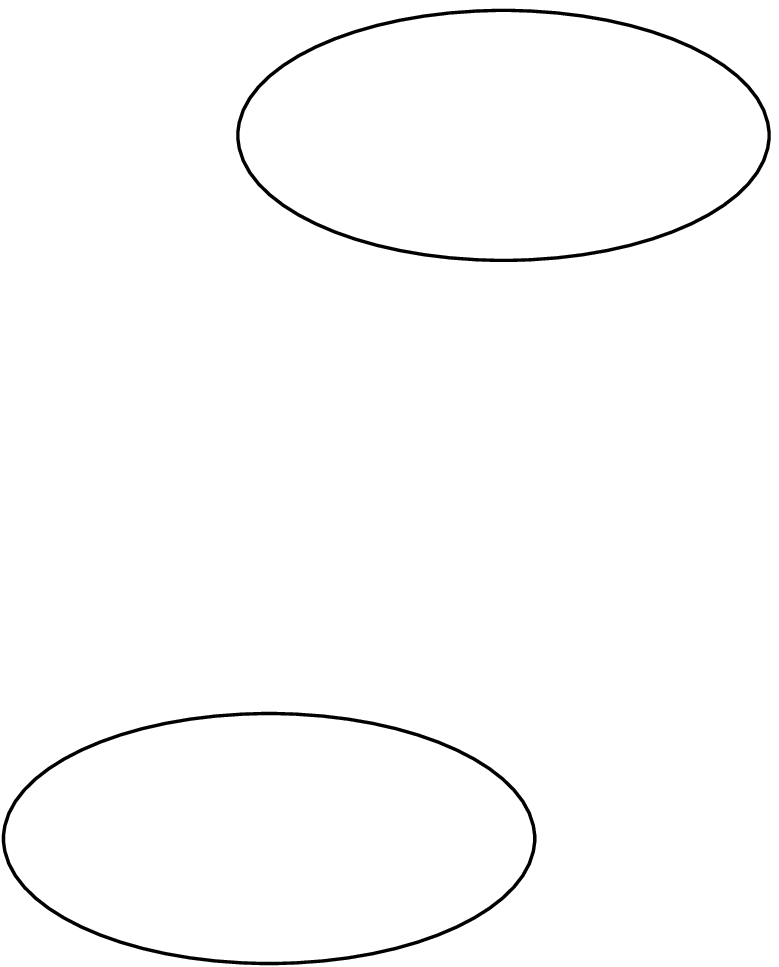,width=5.2cm,height=6.4cm}}
\put(0,10.5){\line(1,3){20}}
\put(45,10.5){\line(1,3){19}}
\put(47,10.5){\line(1,3){8}}
\boldmath
\put(55,36){$v$}
\put(57,40.5){\line(1,3){9}}
\multiput(22,70)(6,-7){6}{\vector(-1,-3){2}}
\multiput(22,50)(5,-4){6}{\vector(-1,-3){2}}
\multiput(22,60)(5,4){6}{\vector(-1,-3){2}}
\multiput(10,20)(5,-4){4}{\vector(-1,-3){2}}
\multiput(35,10)(5,4){2}{\vector(-1,-3){2}}
\multiput(7,30)(2,-4){6}{\vector(-1,-3){2}}
\multiput(17,50)(2,-4){6}{\vector(-1,-3){2}}
\multiput(18,50)(4,4){6}{\vector(-1,-3){2}}
\multiput(40,60)(5,1){4}{\vector(-1,-3){2}}
\multiput(34,50)(2,3){4}{\vector(-1,-3){2}}
\multiput(22,50)(5,-4){5}{\vector(-1,-3){2}}
\end{picture}
\vspace*{1cm}
\caption{
\label{cylinder}
A disc shaped Brownian particle will collide 
with all the molecules with a given velocity $v\/$ in the slanted 
cylinder in unit time.}
\end{figure}

The force on $B\/$ is the momentum transferred to it per
unit time, for which we need to know the number of
collisions per unit time for each value of ${\bf v}\/$.
This number for a given ${\bf v}$ is the number of
 molecules in the (slanted)
cylinder of edge length $\vert {\bf v} \vert$ 
swept in unit time as shown in Figure~\ref{cylinder}, i.e.
\begin{equation}
N_{col} = A \frac{\rho}{m} |v_x|,
\label{noofcols}
\end{equation}
where $A\/$ is the area of cross section of the plate and
$\rho\/$  the density of the gas.  
The force on $B\/$ is
$N_{col}$ times the momentum transfer in each collision,
integrated over all possible velocities:
\begin{equation}
{\bf F} = -A \/\rho\/ (1 + \kappa) 
    \int d {\bf v} P_{\rm c}({\bf v}) |v_x| {\bf v}.
\label{force1}
\end{equation}
where $P_{\rm c}({\bf v})\/$ is the distribution 
of velocities in
the frame of the Brownian particle. The
distribution in the lab frame is Maxwellian i.e.,
\begin{eqnarray}
P(v_x,v_y,v_z) &=& 
P_0 (v_x) P_0 (v_y) P_0 (v_z)
\nonumber \\
P_0 (v_x) 
&=& \frac{1}{\sqrt{2 \pi} {\displaystyle \sigma}} 
{\displaystyle e}^{-\frac{{v_x}^2}{2 {\sigma}^2}}
\label{p0}
\end{eqnarray}
where $\sigma\/$ is the mean thermal velocity component in
one direction.
\begin{equation}
{\sigma}^2 = 
\langle {v_x}^2 \rangle_{\rm th}
=\frac{k_B T}{m}
\label{thermalvelocity}
\end{equation}
If the velocity of a gas molecule is ${\bf v}$ in the
co-moving frame, then it is ${\bf v} + u_0{\hat{\bf x}}$ in
the lab frame.  Therefore the distribution in the co-moving
frame is,
\begin{equation}
P_{\rm c} \; (v_x,v_y,v_z) = 
P_0 (v_x + u_0) P_0 (v_y) P_0 (v_z)
\label{maxwellcomoving}
\end{equation}
The component of the force perpendicular 
to $\hat{x}\/$ adds up to zero 
whereas the $x\/$ component is
\begin{eqnarray}
F_x = A \/\rho\/(1 + \kappa) 
    \int\limits_{-\infty}^{\infty} 
d v_x P_0(v_x + u_0)\/ |v_x|\/ v_x
\nonumber \\
= A \rho (1 + \kappa) 
    \int\limits_{-\infty}^{\infty} 
d v_x P_0(v_x) |v_x-u_0| (v_x - u_0)
\label{force3}
\end{eqnarray}
After a few simple manipulations, we get:
\begin{equation}
F_x = -2 A \rho (1 + \kappa) e^{- \frac{{u_0}^2}{2 {\sigma}^2}}
    \int\limits_0^{\infty} d v_x \/v_x^2
    \frac{1}{\sqrt{2 \pi}}
    e^{- \frac{{v_x}^2}{2 {\sigma}^2}}
    \sinh \left(\frac{v_x u_0}{{\sigma}^2}\right)
\end{equation}
It may be noted that the force is an odd function of
$u_0\/$, in keeping with the general rule that dissipative
terms in an equation of motion violate time reversal
invariance. For small velocities $u_0\/$, we can
Taylor-expand to the lowest order in $u_0\/$ to get   
\begin{equation}
 F_x= -2 A \rho (1 + \kappa) u_0 \sigma
    \frac{1}{\sqrt{2 \pi}}
    \int\limits_0^{\infty} d x 
    x^3 e^{- \frac{{x}^2}{2}}
\label{force5}
\end{equation}
We call the dimensionless integral in the above equation 
$\chi_3\/$ (Appendix~\ref{math}).
The final expression for the force then is:
\begin{equation}
F_x= -2\/ A\/ \rho\/(1 + \kappa)\/ u_0 \sigma {\chi}_3
\label{force6}
\end{equation}
from which we read off the friction coefficient
\begin{equation}
\Gamma = 2 A \rho\/(1 + \kappa)\/ \sigma {\chi}_3
\label{gamma}
\end{equation}
The friction coefficient is not
proportional to the linear dimension of the particle(as it
is in the Stokes' Law) but rather to the area. This result
carries over to the case of the sphere as well. The area
comes  because the gas we are considering is ideal and the
motion of the Brownian particle in no way affects the
 flow of the gas.
The Stokes' result essentially derives from the fact that
the fluid (being a strongly interacting gas) makes way for
the particle as it moves through it~\cite{stokes}.

Let us now proceed with the calculation of the noise.
In the standard Langevin equation~(\ref{langavin-1}) 
the noise $\eta\/$ has to satisfy the fluctuation dissipation
theorem 
\begin{equation}
\langle \eta(t) \eta({\acute t}) \rangle =
 2 k_B T \Gamma \delta ( t - {\acute t})
\label{fdt}
\end{equation}
This theorem relies on the requirement that the Brownian particle
is in thermal equilibrium  with the gas it is moving through.

Before we derive the fluctuation dissipation theorem, we
coarse grain the time in terms of a small time unit
$\epsilon\/$, chosen such that  several collisions take place
within  $\epsilon\/$ time, while the Brownian particle does not
move appreciably during this interval. For a massive enough
particle it is always possible to find such a time. It is
simplest to calculate the noise in the rest frame of the
particle, where the discretized version of the Langevin
equation~(\ref{langavin-1}) is
\begin{equation}
M\/\frac{\Delta v}{\epsilon} = \eta.
\end{equation}
The desirable fluctuation dissipation theorem that the noise
is expected to satisfy then reads in discrete form as
\begin{equation}
\langle {\eta}_i {\eta}_j \rangle
=  2 k_B T \Gamma{\delta}_{ij} / \epsilon
\label{fdtdiscrete}
\end{equation}

\begin{figure}
\unitlength=2mm
\begin{picture}(40,75)(0,0)
\put(0,40){\psfig{figure=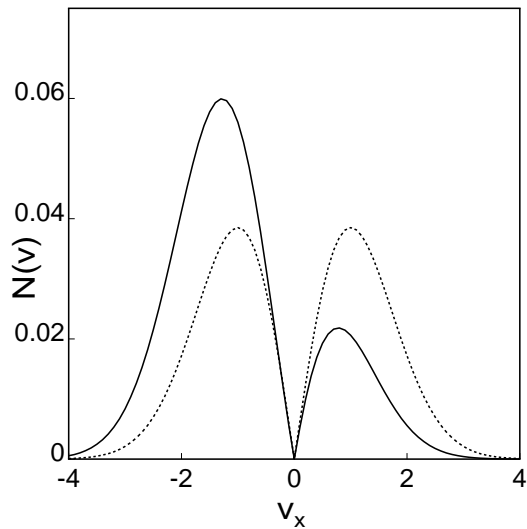,width=8cm,height=8cm}}
\put(0,0){\psfig{figure=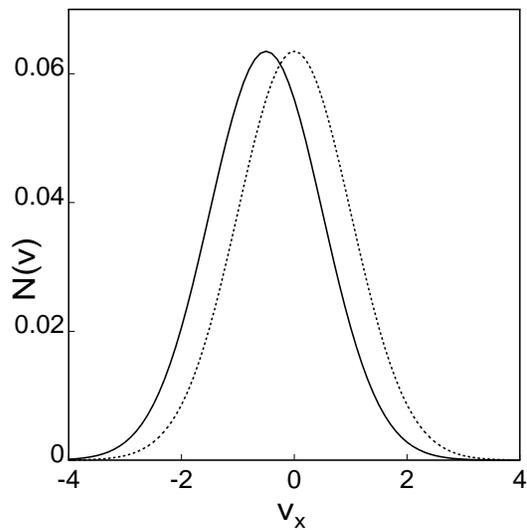,width=8cm,height=8cm}}
\end{picture}
\caption{ \label{plots}
Plot of the distribution $N(v)\/$ for
the ideal gas in two different limits. The upper plot
is for an ideal gas with large mean free path while the
lower one is for a gas with very small mean free path. The dotted 
curves in both the cases are for a Brownian particle at rest and
the solid lines for a moving Brownian particle.}
\end{figure}

Note, that the function $\delta(t_i-t_j)\rightarrow
\delta_{ij}/\epsilon$, such that for $i=j$,
$\delta(0)\rightarrow 1/\epsilon$, which is the appropriate
discrete form \cite{zinn-justin}.
This is the relation we will now derive from the kinematics
of the collisions between the Brownian particle and the gas
molecules. The momentum along $x$, transferred
to B in the interval $\epsilon\/$ is $\epsilon\/\eta$, which
we call $\delta p_x\/$.  We
have to show that:
\begin{itemize}
\item
The momentum transfered $\delta p_x\/$ is distributed
independently in each time interval $\epsilon\/$.
\item
The  mean square value of $\delta p_x\/$ in any given
time is given by
\vspace*{-12pt}
\begin{equation}
\langle (\delta p_x)^2 \rangle = 2 k_B T \Gamma \epsilon 
\label{fdtmomentum}
\end{equation}
\end{itemize}
The first of the above statements  follows from the fact 
that the gas is ideal, so that each collision is independent 
and the coarse graining over time will obviously not generate 
correlations.  
Suppose a total of $N\/$ collisions take place in the interval
$\epsilon\/$, counting all velocities.
The total momentum transferred is~\cite{momentum}
\begin{equation}
\delta p_x = m\/(1 + \kappa) (v_{x1} +  .... +  v_{xN})
\label{momentum1}
\end{equation}
Once again, since the velocities are independent, all cross
terms have a zero average, leading to 
\begin{equation}
\langle {(\delta p_x)}^2 \rangle =
	m^2\/{(1 + \kappa )}^2
	N \langle {v_x }^2 \rangle
\label{momentumsq}
\end{equation}
The number of particles of velocity ${\bf v}\/$ that collide in time
$\epsilon\/$ is
\begin{equation}
N({\bf v}) = \frac{\rho}{m}\/ A\/ \epsilon P_0({\bf v})  |v_x|
\label{cols}
\end{equation}
The total number of collisions is most easily calculated in
Cartesian coordinates:
\begin{eqnarray}
N &=& A \frac{\rho}{m} \epsilon
    \int d {\bf v} P_c ({\bf v})\/ |v_x| \nonumber\\
&=& 2 A \frac{\rho}{m}\/ \sigma\/ \epsilon{\chi}_1 + {\cal
O}((u_0/\sigma)^2) 
\label{totalcols3}
\end{eqnarray}
A calculation in polar coordinates to the leading order gives
\begin{equation}
N =  A \frac{\rho}{m}\/ \sigma\/ \epsilon{\chi}_3
\label{totalcolspolar}
\end{equation}
which is the same result as $\chi_3 /
\chi_1=2\/$ (as shown in Appendix~\ref{math}).  The
average $\langle {v_x}^2 \rangle\/$ is not the thermal
average, but the average over the sample of particles that
collide with B in the interval $\epsilon$.

\begin{eqnarray}
\langle {v_x}^2 \rangle &=&\frac
{\int d {\bf v} N({\bf v}) {v_x }^2} {\int d {\bf v} N({\bf
v}) }
= \frac{1}{N} A \frac{\rho}{m} \epsilon
    \int d {\bf v} P_c ({\bf v})\/ v_x^2 |v_x|\nonumber \\
&=& \frac{1}{2 \sigma {\chi}_1 } 2 {\sigma}^3 {\chi}_3 +
{\cal O}((u_0/\sigma)^2)\nonumber\\
&\simeq& 2 { \sigma}^2
\label{meansqvx5}
\end{eqnarray}
Note that the mean square velocity in Eqn.~(\ref{meansqvx5})
is larger than the thermal average velocity ${\sigma}^2\/$.
This is due to the fact that 
in the ensemble of molecules that collide in a
given interval, there are a greater number of faster molecules, 
as they sweep a larger cylinder (Figure~\ref{plots}).
Substituting for $N\/$ and $\langle v_x^2\rangle\/$ 
in equation~(\ref{momentumsq}) we get
\begin{equation}
\langle {(\delta p_x)}^2 \rangle = m^2\/
	{(1 + \kappa )}^2\/
	2 A \/\frac{\rho}{m}\/ \epsilon\/
        \sigma\/ {\chi}_1\/
	2 {\sigma}^2
\label{psq}
\end{equation}
Combining equations~(\ref{psq}) and~(\ref{gamma}) and using
equations~(\ref{thermalvelocity}) and~(\ref{ratio}) the
above equation becomes
\begin{equation}
\langle {(\delta p_x)}^2 \rangle = (1 + \kappa)\/k_B\/T\Gamma\/\epsilon
\end{equation}
which for $\kappa=1\/$, i.e. for $M >> m\/$ goes to 
the desired fluctuation dissipation result~(\ref{fdtmomentum}).
Also note that the change of momentum (or velocity) of the Brownian
particle, in a time step $\epsilon$, is of the order $\sqrt\epsilon$
and not $\epsilon$, which might be naively expected. This is an essential
feature of Langevin equation \cite{zinn-justin}.
\section{Brownian motion in an Ideal gas with small mean free path}
\label{gas-2}
The classical ideal gas is a familiar concept,  described
by the equation of state
\begin{equation}
PV=nRT.
\label{ideal-eq-st}
\end{equation}
This equation of state is a limiting case of a more
physically reasonable equation of state, namely the real
gas 
(Van der Waals gas) equation~(\cite{real-gas})
\begin{equation}
(P-\frac{a}{V^2})\;(V-b)=nRT.
\label{rgas-eq-st}
\end{equation}
where $a\/$ and $b\/$ represent the intermolecular forces
and the volume occupied by the gas molecules respectively. 
The physical
meaning of working in the ideal gas limit corresponds to
neglecting all forces between the particles other than the
two body hard sphere collisions, and neglecting the volume 
occupied by the molecules themselves as compared to the 
volume of the gas.

It turns out that there is no unique way of reaching such a
limit and different limiting procedures lead to different
physical situations.  One way is to make the gas more and
more dilute in the hope that it will finally become ideal.
As a consequence of dilution, the collision rate is reduced
and finally goes to zero while the mean free path
approaches infinity.  Though such a gas will satisfy the ideal
gas equation of state, it will have infinite relaxation
time, and will never equilibrate on its own! However,
once the equilibrium is established it is maintained in the
absence of external perturbations. This is the kind of gas
we considered in section~\ref{gas-1} for the passage of a Brownian
particle through it.

The second way is more interesting, in that one can consistently
carry out the limiting process from a real gas
to an  ideal
gas while retaining a finite collision rate. In fact it is
possible to fix the collision rate at any desired value and
go through the limiting procedure. As a consequence we can
make the relaxation time as small as we wish.  Therefore, if
disturbed from  equilibrium, such a gas will equilibrate
on its own. 

Both these cases correspond to different physical
situations and the behaviour of a Brownian particle
injected in either of the gases is very different in 
each case.

Going back to the motion of a Brownian particle in second 
type of gas we observe that the relaxation time(the time
in  which the gas goes back to equilibrium, after being
disturbed) can be made smaller than the coarse graining
time (the time for which the Brownian particle does not
change its velocity appreciably). In fact we are allowed to
assume  any preassigned value for the relaxation time and other
parameters can be adjusted accordingly.  The
picture of the previous case where we assumed that the gas is
practically collisionless(at least within coarse graining
time $\epsilon\/$) is no longer valid. The number of
gas particles colliding with the Brownian particle per unit
time at velocity ${\bf v}\/$ is no longer the number of
particles in the slanted cylinder of height $v_x\/$, as the
particles keep colliding and forgetting their velocities within
the time $\epsilon$. 

It turns out that it is simpler to compute the collision
rate in this case as we do not have to worry about the slanted
cylinder. The gas particles colliding with the Brownian
particle are chosen randomly from a continuously
equilibrating velocity distribution, as seen from the rest
frame of the Brownian particle. 
The memory effect present in the slanted cylinder
description of the previous section disappears. 
The number of collisions per unit time $N({\bf v})\/$ 
of the gas molecules moving with a velocity ${\bf v}\/$ with the
Brownian particle in this case (Figure~\ref{plots}) is 
\begin{equation}
N({\bf v}) = K\/ \frac{\rho}{m} A\/ 
v_{\rm th}\/ P_{\rm c}({\bf v}).
\label{col-rate-II}
\end{equation}
where $v_{\rm th}=\sqrt{\frac{\displaystyle k_B T}{\displaystyle
m}\/}$ is the thermal
velocity, K is a normalisation constant 
and all other parameters have the same
meaning as in eq.~(\ref{cols})
The average force can now be readily obtained  as
in the previous case 
\begin{equation}
F=\int N({\bf v}) \delta {\bf p}\/ d{\bf v}.
\end{equation}
Using the value of $\delta {\bf p}\/$ from the 
equation~(\ref{momentum/col})
and of $N({\bf v})\/$ from eqn.~(\ref{col-rate-II}) we have
\begin{eqnarray}
F&=& K\/(1+\kappa)\/ \rho \/ A\/ v_{\rm th} \int P_{\rm c}({\bf
v})\/{\bf v} d{\bf v} 
\nonumber \\
 &=& -u_0 \/ \Gamma 
\end{eqnarray}
where the friction coefficient $\Gamma =
k\/(1+\kappa)\/\rho A\/v_{\rm th}\/$. It is interesting 
that the friction being proportional to the
velocity of the Brownian particle is an exact result here.

We now move on to the noise calculation, the treatment
being very similar to that of section~\ref{gas-1}. 
The momentum transfered
in each time $\epsilon\/$ is obviously independent and we 
basically only have to show that $\langle p^2 \rangle =2 k_B T
\Gamma \epsilon$. The equation~(\ref{momentumsq}) is still
valid for the Brownian plate where only the $x\/$ component of 
momentum is important but the expressions for the total number of collisions 
$N\/$ and $\langle v_x^2 \rangle \/$ are different in
this case. We have for the Brownian particle at rest 
\begin{eqnarray}
N &=& \int N({\bf v}) d{\bf v} 
\nonumber \\
  &=& K\/ \frac{\rho}{m} A v_{th}
\end{eqnarray}
and since there are no memory effects we have the pure
thermal velocity distribution
with
\begin{equation}
\langle v^2_x \rangle = \frac{k_B T}{m}
\end{equation}
Using these expressions and the equation~(\ref{momentumsq})
we get 
\begin{eqnarray}
\langle p_x^2 \rangle &=& m^2\/(1+\kappa)^2 K \frac{\rho}{m}
 A v_{\rm th} \epsilon \frac{k_B T}{m}
\nonumber \\
&=& (1+\kappa)\/\epsilon k_B T\/\/ \Gamma
\end{eqnarray}
In the limit when Brownian particle is of large mass
compared to the gas molecules i.e. $\kappa \rightarrow 1\/$
we recover the fluctuation dissipation
theorem~(\ref{fdtmomentum}). The derivation has been
simpler and more exact in this case, where the gas 
is constantly equilibrating. We have not used any
approximations here as opposed to Section~II. It is also clear in this
case that the separation of time scales for the motion of the
gas molecules and that of  the Brownian particle is essential for
the fluctuation dissipation theorem to hold and thus the theorem
is valid only in the limit of $\kappa \rightarrow 1$.
\section{Concluding remarks}
\label{conc}
We have shown,
how the familiar result of Langevin equation and
fluctuation dissipation theorem can be derived and
understood in a special case when the fluid through which
the Brownian particle moves is an ideal gas. The frictional
force comes about because of the fact that if $B\/$
moves in a particular direction then as seen from the frame
of $B\/$ the velocity of gas molecules hitting it from
the front is more compared to that for the molecules
hitting it from behind. As the details of the calculations
in section~II and~III show, the situation is a little 
more subtle and we have to be careful while setting up the
expression for the collision rate as a function of velocity.
Moreover, the separation of time scales for the motion of 
the Brownian particle and the gas particles is essential
for obtaining the fluctuation dissipation theorem. 

As a matter of contrast we observe that the essential
difference in the two kind of ideal gases is in the
distribution $N({\bf v})\/$ of collision rate is  at velocity 
${\bf v}\/$.
For the ideal gas of the first kind, this function tends to 
have larger values for larger velocities as compared to the
Maxwellian, as the volume of the cylinder from which all
the molecules must collide with $B\/$ is directly
proportional to $|{\bf v}|\/$. On the other hand, for the
ideal gas of the second type, the absence of memory effects
and the breakdown of the cylinder picture prevents any such
biases.

Another interesting observation is that the friction
coefficient is proportional to the cross-section area
for both the cases of the plate and the sphere unlike in 
the case of Stokes' law where it is proportional to the radius. 
This is due to the fact that we have
neglected the back reaction due to the motion of the sphere
on the velocity distribution of the gas particles, which
has been retained to be Maxwellian all along. On the
contrary for the calculation of the Stokes' law one takes
into account the effect of the motion of the sphere on the
fluid velocity.  While our treatment holds for a dilute
noninteracting gas, the Stokes' law is obeyed in fluids
with higher density.

We feel that our analysis can be carried out for other kind
of Brownian motion as well. One could, for example, consider
the Brownian motion of angular variables for molecules
moving in an ideal gas . We expect that interesting physical
insights can be gained if such an analysis is carried out.

\appendix
\section{Calculation of friction coefficient for a Brownian sphere} 
\label{sphere}
In this appendix, we demonstrate that the friction
coefficient can be calculated for the case when the
Brownian particle is a sphere instead of a plate. The
underlying physics is same though  the calculation is a little more
involved. The plate was chosen
only for simplicity and this calculation is given here to
convince the reader that there is nothing special about the
plate.  We will see that the only difference from the plate
case is a geometrical factor.

\begin{figure}
\unitlength=1mm
\begin{picture}(80,80)(0,0)
\boldmath
\put(0,0){\psfig{figure=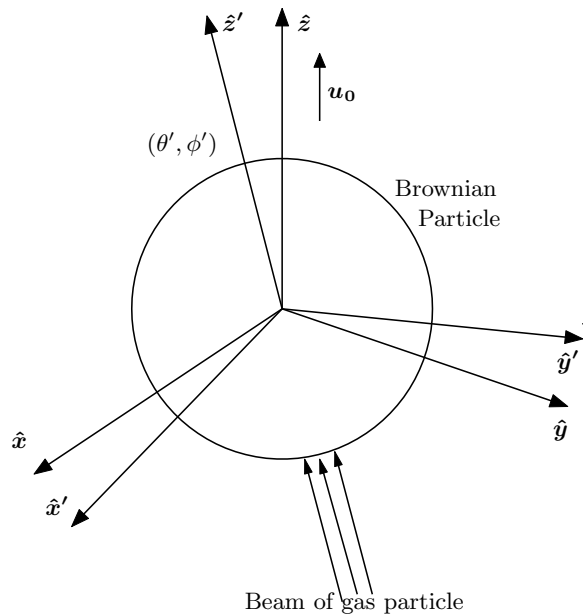,width=8cm,height=8cm}}
\put(76,23){$\hat{y}$}
\put(76.5,32){$\hat{y}^{\prime}$}
\put(42,77){$\hat{z}$}
\put(32,77){$\hat{z}^{\prime}$}
\put(4,21.5){$\hat{x}$}
\put(8.5,12.5){$\hat{x}^{\prime}$}
\put(46,68){$u_0$}
\unboldmath
\put(22,61){$(\theta^{\prime},\phi^{\prime})$}
\put(35,0){Beam of gas particle}
\put(55,55){Brownian}
\put(57,51){ Particle}
\end{picture}
\vspace*{1cm}
\caption{ \label{sphere-pic} A pictorial description of collisions for a 
spherical Brownian particle.}
\end{figure}

Imagine the Brownian particle to be a sphere of radius
$R\/$ moving in the $\hat{\bf z}$ direction with velocity
$u_o$(As shown in Figure~\ref{sphere-pic}). 
We consider a frame of reference $F\/$ which is at
rest with respect to the Brownian particle.  Consider the
collision of the Brownian particle with a beam of particles
moving with velocity ${\bf V}(V_o,\theta',\phi')\/$ in this
frame. Further consider a frame $F^{\prime}\/$ again at
rest with respect to the Brownian particle but with its
orientation such that the $\hat{\bf z^{\prime}}\/$ axis is in
the direction of $\bf{V}\/$. When a narrow beam hits the
sphere at a point ($R,\theta,\phi\/$) in the frame
$F^{\prime}\/$, it is reflected by the sphere and the force
exerted on the sphere, which is  along ${\bf
z}^{\prime}\/$ direction is given by
\begin{eqnarray}
{\bf F}_{\hat{\bf z^{\prime}}-{\rm rest}} 
&=& {\rm flux}.(1+\kappa).V_o\/\cos\theta
\;  \hat{\bf z}^{\prime}.
\nonumber \\
{\rm Where}&&  \nonumber \\ 
 {\rm flux} &=& \rho\/{\bf V}. (-d{\bf s}) 
= \rho\/V_o\/\cos\theta
\/R^{2}\/\sin\theta\/ d\theta\/ d\phi 
\label{eq-force}
\end{eqnarray}

The velocity distribution of
equation~(\ref{maxwellcomoving}) breaks the $\theta'$
symmetry but preserves the $\phi'$ symmetry, therefore in
the frame $F\/$ the $\hat{\bf x}\/$ and $\hat{\bf y}\/$
component of the force averages to zero and we need to 
consider only the $\hat{\bf z}\/$ component of the force.
\begin{equation}
F_{\hat{\bf z}} = 
F_{\hat{\bf z}^{\prime}-{\rm rest}}.\cos\theta^{\prime}.
\end{equation}
Rewriting the eqn~(\ref{maxwellcomoving}) for the velocity
distribution in the spherical polar coordinates as seen
from the frame $F\/$ we have
\begin{equation}
P_{\rm c}({\bf V}) = \frac{1}{(\sqrt{2\pi}\sigma)^3} e^{
{{-[(V_o\cos\theta^{\prime} + u_o)^2 + (V_o\sin\theta^{\prime})}^2
]}/{2\sigma^2} }
\end{equation}

The total frictional force can now be readily obtained by 
integrating over all the angular variables and the velocity
distribution
\begin{eqnarray}
F_{\hat{Z}} =&&
C_1\/C_2\/\int_{0}^{\infty}dV_{o}V_{o}^{4}e^{-V_{o}^2/{2\sigma^2}}
\nonumber \\
&&\quad\int_{0}^{\pi}\frac {1} { ({\sqrt{2\pi}\sigma})^3 } 
e^{2u_{o}V_{o}\cos\theta'/{2\sigma^2}}\/
\sin\theta'\cos\theta'd{\theta'}
\end{eqnarray}
where, 
\begin{equation}
C_1 = {(2\pi)^2}(1+\kappa){R^2}
{\displaystyle e}^{ \frac{-u_o^2}{2\sigma^2} }
\; {\rm and} 
\;C_2 = \int_{0}^{\pi}d \theta \sin\theta\cos^{2}\theta = 2/3
\end{equation}
To the lowest order, the $\int d \theta^{\prime}\/$ gives
$u_{o}V_{o}/{3{\sigma^2}}\/$ and by working out the 
$\int dV_o\/$ we obtain the total frictional force
\begin{equation}
F_{\hat{Z}} = -C_1\/C_2\/\frac{u_{o}\sigma}{6\pi}{\chi_5}
\end{equation}

Using the dimensionless integral $\chi_5$ from
appendix~\ref{math},
the relation $\chi_5=4\chi_3\/$ 
and substituting the values of $C_1,C_2$, the
final expression for the force simplifies to
\begin{equation}
F_{\hat{Z}} = -\frac{8}{9}\/\left(2A\/ \rho(1 +
\kappa)\/{u_{o}\/\sigma}\/\chi 3 \right)
\end{equation}
where $A=\pi R^2\/$ is the cross section of the sphere.
We notice that the only difference in this expression as
compared to the plate case of eqn.~(\ref{force6}) is
the geometrical factor of $8/9 \/$.
\section{Some useful Integrals}
\label{math}
We list  some useful integrals used in the paper.
\begin{eqnarray}
\Gamma(n) &=& \int\limits_{0}^{\infty}
dx \,x^{n-1} \, e^{-x}
 = 2 \/ \int\limits_{0}^{\infty}
dx \,x^{2n-1} \, e^{-{x^2}} \nonumber\\
&=& 2^{1-n} \/ \int\limits_{0}^{\infty}
dx \,x^{2n-1} \, e^{-\frac{x^2}{2}}
\\
\chi _n &=& \frac{1}{\sqrt{2 \pi}} \/ \int\limits_{0}^{\infty}
dx \,x^{n} \, e^{-\frac{x^2}{2}}
= \frac{1}{\sqrt{2 \pi}} \/ 
	2^{\frac{n-1}{2}} \,\,
	\Gamma \left( \frac{n+1}{2}\right) 
\nonumber
\\
&&\chi_3/\chi_1=2
\quad
\chi_5/\chi_3=4
\label{ratio}
\end{eqnarray}


\begin{thebibliography}{99}
\bibitem{momentum}
In fact the number of collisions in the interval is
itself a fluctuating quantity, however, in our calculation 
we assume it to be a constant. This is a good approximation
as long as $N\/$ is large, as the fluctuation in $N\/$ is
$O(\sqrt{N})\/$, which is much smaller than $N$. Since
$N\/$ is proportional to the size of the Brownian particle
as well as to the interval $\epsilon\/$, it is always large
for a large particle. Note that a more massive particle
moves less in a given interval, so $\epsilon\/$ can be chosen
to be large. Hence both the mass and the size of the
particle contribute to increasing $N\/$.
\end{thebibliography}
\end{document}